
%
%
%
%
%
\magnification=1200
%
%
\hsize=31pc
\vsize=55 truepc
\hfuzz=2pt
\vfuzz=4pt
\pretolerance=5000
\tolerance=5000
\parskip=0pt plus 1pt
\parindent=16pt
%
%
\font\fourteenrm=cmr10 scaled \magstep2
\font\fourteeni=cmmi10 scaled \magstep2
\font\fourteenbf=cmbx10 scaled \magstep2
\font\fourteenit=cmti10 scaled \magstep2
\font\fourteensy=cmsy10 scaled \magstep2
\font\large=cmbx10 scaled \magstep1
%

%
%

%
%

%
%

%
%
\font\eightrm=cmr8
\font\eighti=cmmi8
\font\eightbf=cmbx8
\font\eightit=cmti8

\font\eightsy=cmsy8
\font\sixrm=cmr6
\font\sixi=cmmi6
\font\sixsy=cmsy6

\def\tenpoint{\def\rm{\fam0\tenrm}%
  \textfont0=\tenrm \scriptfont0=\sevenrm
                      \scriptscriptfont0=\fiverm
  \textfont1=\teni  \scriptfont1=\seveni
                      \scriptscriptfont1=\fivei
  \textfont2=\tensy \scriptfont2=\sevensy
                      \scriptscriptfont2=\fivesy
  \textfont3=\tenex   \scriptfont3=\tenex
                      \scriptscriptfont3=\tenex
  \textfont\itfam=\tenit  \def\it{\fam\itfam\tenit}%
  \textfont\slfam=\tensl  \def\sl{\fam\slfam\tensl}%
  \textfont\bffam=\tenbf  \scriptfont\bffam=\sevenbf
                            \scriptscriptfont\bffam=\fivebf
                            \def\bf{\fam\bffam\tenbf}%
  \normalbaselineskip=20 truept
  \setbox\strutbox=\hbox{\vrule height14pt depth6pt
width0pt}%
  \let\sc=\eightrm \normalbaselines\rm}
\def\eightpoint{\def\rm{\fam0\eightrm}%
  \textfont0=\eightrm \scriptfont0=\sixrm
                      \scriptscriptfont0=\fiverm
  \textfont1=\eighti  \scriptfont1=\sixi
                      \scriptscriptfont1=\fivei
  \textfont2=\eightsy \scriptfont2=\sixsy
                      \scriptscriptfont2=\fivesy
  \textfont3=\tenex   \scriptfont3=\tenex
                      \scriptscriptfont3=\tenex
  \textfont\itfam=\eightit  \def\it{\fam\itfam\eightit}%
  \textfont\bffam=\eightbf  \def\bf{\fam\bffam\eightbf}%
  \normalbaselineskip=16 truept
  \setbox\strutbox=\hbox{\vrule height11pt depth5pt width0pt}}
\def\fourteenpoint{\def\rm{\fam0\fourteenrm}%
  \textfont0=\fourteenrm \scriptfont0=\tenrm
                      \scriptscriptfont0=\eightrm
  \textfont1=\fourteeni  \scriptfont1=\teni
                      \scriptscriptfont1=\eighti
  \textfont2=\fourteensy \scriptfont2=\tensy
                      \scriptscriptfont2=\eightsy
  \textfont3=\tenex   \scriptfont3=\tenex
                      \scriptscriptfont3=\tenex
  \textfont\itfam=\fourteenit  \def\it{\fam\itfam\fourteenit}%
  \textfont\bffam=\fourteenbf  \scriptfont\bffam=\tenbf
                             \scriptscriptfont\bffam=\eightbf
                             \def\bf{\fam\bffam\fourteenbf}%
  \normalbaselineskip=24 truept
  \setbox\strutbox=\hbox{\vrule height17pt depth7pt width0pt}%
  \let\sc=\tenrm \normalbaselines\rm}

\def\today{\number\day\ \ifcase\month\or
  January\or February\or March\or April\or May\or June\or
  July\or August\or September\or October\or November\or
December\fi
  \space \number\year}
%
%
\newcount\secno      
\newcount\subno      
\newcount\subsubno   
\newcount\appno      
\newcount\tableno    
\newcount\figureno   
\normalbaselineskip=20 truept
\baselineskip=20 truept
%
%
\def\title#1
   {\vglue1truein
   {\baselineskip=24 truept
    \pretolerance=10000
    \raggedright
    \noindent \fourteenpoint\bf #1\par}
    \vskip1truein minus36pt}
%
%
\def\author#1
  {{\pretolerance=10000
    \raggedright
    \noindent {\large #1}\par}}
%
%
\def\address#1
   {\bigskip
    \noindent \rm #1\par}
%
%
\def\shorttitle#1
   {\vfill
    \noindent \rm Short title: {\sl #1}\par
    \medskip}
%
%
\def\pacs#1
   {\noindent \rm PACS number(s): #1\par
    \medskip}
%
%
\def\jnl#1
   {\noindent \rm Submitted to: {\sl #1}\par
    \medskip}
%
%
\def\date
   {\noindent Date: \today\par
    \medskip}
%
%

%
%
\def\keyword#1
   {\bigskip
    \noindent {\bf Keyword abstract: }\rm#1}
%
%

%
%
%

%
%
\def\entry#1#2#3
   {\noindent
    \hangindent=20pt
    \hangafter=1
    \hbox to20pt{#1 \hss}#2\hfill #3\par}
%
%
\def\subentry#1#2#3
   {\noindent
    \hangindent=40pt
    \hangafter=1
    \hskip20pt\hbox to20pt{#1 \hss}#2\hfill #3\par}
%
%
\def\section#1
   {\vskip0pt plus.1\vsize\penalty-250
    \vskip0pt plus-.1\vsize\vskip24pt plus12pt minus6pt
    \subno=0 \subsubno=0
    \global\advance\secno by 1
    \noindent {\bf \the\secno. #1\par}
    \bigskip
    \noindent}
%
%
\def\subsection#1
   {\vskip-\lastskip
    \vskip24pt plus12pt minus6pt
    \bigbreak
    \global\advance\subno by 1
    \subsubno=0
    \noindent {\sl \the\secno.\the\subno. #1\par}
    \nobreak
    \medskip
    \noindent}
%
%
\def\subsubsection#1
   {\vskip-\lastskip
    \vskip20pt plus6pt minus6pt
    \bigbreak
    \global\advance\subsubno by 1
    \noindent {\sl \the\secno.\the\subno.\the\subsubno. #1}\null. }
%
%
\def\appendix#1
   {\vskip0pt plus.1\vsize\penalty-250
    \vskip0pt plus-.1\vsize\vskip24pt plus12pt minus6pt
    \subno=0
    \global\advance\appno by 1
    \noindent {\bf Appendix \the\appno. #1\par}
    \bigskip
    \noindent}
%
%
\def\subappendix#1
   {\vskip-\lastskip
    \vskip36pt plus12pt minus12pt
    \bigbreak
    \global\advance\subno by 1
    \noindent {\sl \the\appno.\the\subno. #1\par}
    \nobreak
    \medskip
    \noindent}
%
%
\def\ack
   {\vskip-\lastskip
    \vskip36pt plus12pt minus12pt
    \bigbreak
    \noindent{\bf Acknowledgments\par}
    \nobreak
    \bigskip
    \noindent}
%
%

%
%
\def\tabcaption#1
   {\global\advance\tableno by 1
    \noindent {\bf Table \the\tableno.} \rm#1\par
    \bigskip}
%
%

%
%
\def\figcaption#1
   {\global\advance\figureno by 1
    \noindent {\bf Figure \the\figureno.} \rm#1\par
    \bigskip}
%
%
\def\references
     {\vfill\eject
     {\noindent \bf References\par}
      \parindent=0pt
      \bigskip}
%
%
\def\refjl#1#2#3#4
   {\hangindent=16pt
    \hangafter=1
    \rm #1
   {\frenchspacing\sl #2
    \bf #3}
    #4\par}
%
%
\def\refbk#1#2#3
   {\hangindent=16pt
    \hangafter=1
    \rm #1
   {\frenchspacing\sl #2}
    #3\par}
%
%
\def\numrefjl#1#2#3#4#5
   {\parindent=40pt
    \hang
    \noindent
    \rm {\hbox to 30truept{\hss #1\quad}}#2
   {\frenchspacing\sl #3\/
    \bf #4}
    #5\par\parindent=16pt}
%
%
\def\numrefbk#1#2#3#4
   {\parindent=40pt
    \hang
    \noindent
    \rm {\hbox to 30truept{\hss #1\quad}}#2
   {\frenchspacing\sl #3\/}
    #4\par\parindent=16pt}
%
%
\def\dash{---{}---}
%
%
\def\frac#1#2{{#1 \over #2}}
%
%

%
%
\def\d{{\rm d}}
%
%
\def\e{{\rm e}}
%
%
\def\i{\ifmmode{\rm i}\else\char"10\fi}
%
%

%
%

%
%

%
%

%
%
\catcode`\@=11
%
%
\def\ind{\hbox to 5pc{}}
%
%
\def\eq(#1){\hfill\llap{(#1)}}
%
%

%
%
\def\deqn#1{\displ@y\halign{\hbox to \displaywidth
    {$\@lign\displaystyle##\hfil$}\crcr #1\crcr}}
%
%
\def\indeqn#1{\displ@y\halign{\hbox to \displaywidth
    {$\ind\@lign\displaystyle##\hfil$}\crcr #1\crcr}}
%
%
\def\indalign#1{\displ@y \tabskip=0pt
  \halign to\displaywidth{\ind$\@lign\displaystyle{##}$\tabskip=0pt
    &$\@lign\displaystyle{{}##}$\hfill\tabskip=\centering
    &\llap{$\@lign##$}\tabskip=0pt\crcr
    #1\crcr}}
\catcode`\@=12
%
%



%
%

\def\JPA{J. Phys. A: Math. Gen.}


%
%

\def\PL{Phys. Lett.}

%
%

\def\one#1{#1^{\raise4pt\hbox{$\scriptstyle\!\!\!\!1$}}\,{}}
\def\onepar#1{#1^{\raise4pt\hbox{$\scriptstyle\!\!\!\!(1)$}}\,{}}
\def\two#1{#1^{\raise4pt\hbox{$\scriptstyle\!\!\!\!2$}}\,{}}
\def\twopar#1{#1^{\raise4pt\hbox{$\scriptstyle\!\!\!\!(2)$}}\,{}}
\def\twoprime#1{#1^{\raise4pt\hbox{$\scriptstyle\!\!\!\!{2^\prime}$}}\,{}}
\def\three#1{#1^{\raise4pt\hbox{$\scriptstyle\!\!\!\!3$}}\,{}}
\def\om{\omega}
\def\a{\alpha}
\def\b{\beta}
\def\g{\gamma}
\def\d{\delta}
\def\A{{\cal A}}
\def\F{F}
\def\eps{\varepsilon}
\def\id{{\rm id}}
\def\D{\Delta}
\def\C{{\bf C}}
\def\R{{\bf R}}
\def\phi{\varphi}
\def\P{{\cal P}}
\def\qdet{{\rm det}_q}
\def\qtr{{\rm tr}_q}
\def\tr{{\rm tr}}
\def\B{{\cal B}}
\def\Uq{{\cal U}_q(sl(2))}
\def\X{{\cal X}}
\def\FR{F_R}
\def\AR{{\cal A}_R}
\def\cf{{\it cf.\ }}
\def\tilde{\widetilde}

{\bf Yukawa Institute Kyoto} \hfill YITP/K-980

\hfill May 1992

\title{Algebraic structures related to reflection equations}

\author{P P Kulish\dag\footnote\ddag{\rm Permanent address:
Steklov Mathematical Institute, Fontanka 27,
St.Petersburg 191011, Russia}
and E K Sklyanin\S\ddag}
\address{\dag\ Yukawa Institute for Theoretical Physics (YITP),
Kyoto University, Kyoto 606, Japan}
\address{\S\ Research Institute for Mathematical Sciences (RIMS),
Kyoto University,  Kyoto 606, Japan}
\bigskip
\bigskip
\noindent{\bf Abstract}
\bigskip

  Quadratic algebras related to the reflection equations are introduced.
They are quantum group comodule algebras. The quantum group $F_q(GL(2))$
is taken as the example. The properties of the algebras (center,
representations, realizations, real forms, fusion procedure etc) as well
as the generalizations are discussed.

\section{Introduction}

The progress in understanding the algebraic roots of the quantum
integrability achieved in last decades has already resulted in introducing
several new algebraic objects, such as the Yang-Baxter equation (YBE),
quantum groups and quantum algebras, exchange and quadratic algebras.
It seems that the list should be enriched with another item: the reflection
equation (RE), which has arisen recently in several independent contexts.

The RE reads
$$ R(u-v)\one K(u)R(u+v)\two K(v)=
   \two K(v)R(u+v)\one K(u)R(u-v)  \eqno(1)
$$
where $K(u)$ is a square $N\times N$ matrix,
$\one K\equiv K\otimes\id _N$,
$\two K\equiv \id _N\otimes K$
(the notation is usual in the quantum
inverse scattering method), and $R(u)$ is a solution to the YBE.
One can also consider (1) as defining relations for the associative algebra
$\A$ generated by the elements of the matrix $K(u)$ [1].

Being introduced in [2] as the equation describing factorized
scattering on a half-line, the RE and the related algebra $\A$ had found soon
quite different applications to the quantum current algebras [3] and to the
integrable models with non-periodic boundary conditions [1, 4].

Since the constant ({\it i.e.} not including the spectral parameter $u$)
solutions
to the YBE are extensively used in the quantum group theory [5], it is quite
natural to study a version of the RE without spectral parameter. Though
such an equation
and the related algebras (for different R matrices) appeared already in
several papers [5--12], they were not distinguished as separate
objects of study until recently.

In the present paper we collect the basic facts  which
are necessary for any systematic study of the RE. We restrict ourselves
to the case of the simplest quantum group $F_q(GL(2))$ and choose the following
form of RE
$$ R\one  KR^{t_1}\two K=
   \two K R^{t_1}\one K R              \eqno(2)
$$

The paper is organized as follows. After introducing the basic definitions
and notation the general properties of the quadratic algebra $\A$ defined by
(2) are described in section 2. They are discussed and partially proved in
section 3. In section 4 few comments are given on the representations of $\A$.
In section 5 some equivalent  variants of RE are pointed out
as well as the relations between $\A$ and  other algebras.
In conclusion, we discuss some generalizations
of RE and its applications.

\section{Definitions}

The quantum group $\F\equiv F_q(GL(2))$ can be defined  as the associative
algebra
generated by  four elements $a$, $b$, $c$, $d$ and the relations
$$ \eqalign{ab&=qba,\cr ac&=qca,} \qquad
   \eqalign{bd&=qdb,\cr cd&=qdc,}  \qquad
   \eqalign{[a,d]&=\om bc,\cr [b,c]&=0,}       \eqno(3)
$$
$q$ being a complex parameter and $\om=q-q^{-1}$. Introducing matrix
$T$
$$ T=\pmatrix{a&b\cr c&d} $$
and using the notation $\one T\equiv T\otimes\id$,
$\two T\equiv \id\otimes T$
one can rewrite the relations (3)  in the compact form [5]
$$ R\one T\two T=\two T\one T R          \eqno(4) $$
where the R matrix is given by
$$ R=\pmatrix{q&&&\cr &1&&\cr &\om&1&\cr &&&q}, \qquad
   R^{t_1}=\pmatrix{q&&&\om\cr &1&&\cr &&1&\cr &&&q}     \eqno(5)
$$
and $t_1$ means transposition with respect to the first space in $\C^2
\otimes\C^2$.

{\it Remark.} If we replace the R matrix in (4) by $\P R^{-1}\P$,
where $\P$ is the permutation operator in $\C^2\otimes\C^2$, the commutation
relations (3) remain the same. The consequences of this fact are discussed
in section 6.2.

The Hopf algebra structure is defined on $\F$ after introducing
 comultiplication map $\D:\F\rightarrow \F\otimes \F$, counit map
$\eps:\F\rightarrow\C$  and coinverse map $s:\F\rightarrow \F$ by the
formulae
$$ \D(T)=T_1T_2, \qquad
   \eps(T)=\id , \qquad
   s(T)=T^{-1}
$$
where $(T_1T_2)_{ij}\equiv \sum_{k=1}^2 T_{ik}\otimes T_{kj}$.

Now define the associative algebra $\A$ by four generators $\a$, $\b$,
$\g$, $\d$, which can be thought of as the elements of the matrix $K$
$$ K=\pmatrix{\a&\b\cr \g&\d},                             $$
and by the quadratic relations (2) or, more explicitely,
$$ \eqalign{[\a,\b]&=\om\a\g,\cr [\b,\g]&=0,} \qquad
   \eqalign{\a\g&=q^2\g\a,\cr [\b,\d]&=\om\g\d,} \qquad
   \eqalign{[\a,\d]&=\om(q\b+\g)\g,\cr \g\d&=q^2\d\g.}        \eqno(6)
$$

The algebra $\A$ has the following properties.

1. $\A$ is a Poincar\'e-Birkhoff-Witt (PBW) algebra which means that the linear
space spanned by the monomials of order $p$ in the generators $\a\b\g\d$ has
the same dimension as in the commutative case that is $(p+3)!/p!3!$

2. $\A$ is an $\F$-comodule-algebra  that is there exists a map $\phi$
(coaction of $\F$) $\phi:\A\rightarrow \F\otimes\A$ which is consistent
with the comultiplication $\D$
$$ (\D\otimes\id)\circ\phi=(\id\otimes\phi)\circ\phi              \eqno(7)$$
and the counit $\eps$
$$ (\eps\otimes\id)\circ\phi=\id                                  \eqno(8)$$
and, besides, is an algebra homomorphism.

By virtue of the duality between $\F=F_q(GL(2))$ and the quantum algebra $\Uq$
the dual map $\phi^*:\Uq\otimes\A\rightarrow \A$ defines the structure
of $\Uq$-module algebra on $\A$.

3. The center of $\A$ is generated by two elements (for generic $q$,
not root of unity)
$$ c_1=\b-q\g, \qquad c_2=\a\d-q^2\b\g                    \eqno(9) $$

4. There are three real forms of $\A$ consistent with three known real forms
of $\F$: $F_q(U(2))$, $F_q(U(1,1))$ and $F_q(GL(2,\R))$.

5. The two-sided ideal generated in $\A$ by the relation $c_1=0$ is invariant
under
the coaction $\phi$ and the corresponding quotient algebra
is isomorphic to the quantum homogeneous space.

\section{Discussion and Proofs}

1. In order to prove the PBW-property of $\A$ it is necessary to verify that
any monomial in $\a\b\g\d$
can be expanded uniquely into the sum of alphabetically ordered monomials.
The possibility of alphabetical ordering is easy to establish using the
commutation relations (6). As for the linear independence of alphabetically
ordered monomials, it  is sufficient to verify it only for cubic monomials
[13, 14].

2. The (left) coaction of $\F$ on $\A$ is defined on the generators by the
formula
$$ \phi(K)=TKT^t                                           \eqno(10)$$
where $t$ stands for the matrix transposition, and is extended to the whole
algebra $\A$ as an algebra homomorphism. For example,
$$ \phi(\b)=ac\a+bc\g+ad\b+bd\d                              $$
Verification of the
$\F$-comodule-algebra axioms (7), (8) is a matter of direct calculation
based on the commutation relation (4) and the equivalent relations
$$ R\one T^t\two T^t=\two T^t\one T^tR, \qquad
   \one T^tR^{t_1}\two T=\two TR^{t_1}\one T^t, \qquad
   \one TR^{t_1}\two T^t=\two T^tR^{t_1}\one T                $$
obtained from (4) using transpositions and the symmetries
$$ \P R^t\P=R, \qquad [\P,R^{t_1}]=0          \eqno(11) $$
of the R matrix (5).
Here superscript $t$ means total transposition and $\P$ is the permutation
operator in $\C^2\otimes\C^2$.

The duality between the quantum group $\F=F_q(GL(2))$ and the quantum
algebra $\Uq$ [5, 15] implies that $\A$ is also a $\Uq$-module algebra.

The algebra $\Uq$ is generated by three elements $H$, $X_+$, $X_-$ and the
relations
$$ [H,X_\pm]=\pm X_\pm, \qquad [X_+,X_-]={q^{2H}-q^{-2H} \over q-q^{-1}}
                                                                        $$
or in the matrix form
$$ R^\pm \one L^\pm \two L^\eps = \two L^\eps \one L^\pm R^\pm
   \qquad \forall \eps\in\{+,-\}                                       $$
where
$$ L^+=\pmatrix{q^H & \om X_- \cr 0 & q^{-H} }, \qquad
   L^-=\pmatrix{q^{-H} & 0 \cr -\om X_+ & q^H }                    \eqno(12)$$
and
$$ R^+=q^{-1/2}\P R\P, \qquad R^-=q^{1/2}R^{-1}                           $$

The pairing $\left<\,,\,\right>$ between $\F$ and $\Uq$ is described by the
relations [5]
$$ \left<\one L^\pm,\two T\right>=R_{12}^\pm,\qquad
   \left<\one L^\pm,\two T\three T\right>=R_{12}^\pm R_{13}^\pm,\ldots
                                                                        $$
where, as usual, the subscripts mark the spaces where the corresponding
R matrices act nontrivially.

The above formulas allow to calculate
the corresponding action $\phi^*$ of $\Uq$ on $\A$:
$$ \phi^*(\one L^\pm)\two K\rightarrow R_{12}^\pm\two K(R_{12}^\pm)^{t_2}
                                                                         $$
or, more explicitely,
$$ \eqalign{\phi^*(X_-) & \pmatrix{\a&\b\cr \g&\d}=
              \pmatrix{0&\a\cr q^{-1}\a&\b+q^{-1}\g} \cr
 \phi^*(X_+) & \pmatrix{\a&\b\cr \g&\d}=
               \pmatrix{q\b+\g & q\d \cr \d & 0} \cr
 \phi^*(q^H) & \pmatrix{\a&\b\cr \g&\d}=
               \pmatrix{q\a & \b \cr \g & q^{-1}\d} }               $$

3. Before we proceed to the discussion of the center of $\A$ let us recall few
facts about the quantum group $\F$.
The quantum determinant $\qdet T$ generating the center of $\F$ can be
constructed via the fusion procedure technique [5, 16]
$$ P_-\one T\two T=P_-\one T\two TP_-=\qdet TP_-                \eqno(13)$$

The rank one projector $P_-$ and the complementary rank three projector
$P_+$ in $\C^2\otimes\C^2$ are defined by the spectral decomposition
of the modified R matrix [17]
$$ \hat R\equiv\P R=qP_+-q^{-1}P_-                           \eqno(14)$$

The commutativity of $\qdet T$ with $T$ and its group-like property
$\D(\qdet T)=\qdet T\otimes\qdet T$ follow from the relation (4), e.g.
$$ \D(\qdet T)=P_-\one T_1\one T_2\two T_1\two T_2=
               P_-\one T_1\two T_1P_-\one T_2\two T_2=
               \qdet T\otimes \qdet T                          $$

The quantum group $F_q(GL(2))$ is well known to possess the invariant
quadratic form $\eps_q$
$$ \eps_q=\pmatrix{0&1\cr -q&0}, \qquad
   T\eps_qT^t=T^t\eps_qT=\eps_q\qdet T                         \eqno(15)$$

The last relation, rewritten as $\sum_{mn}(\eps^t_q)_{mn}T_{ni}T_{jm}=
(\eps^t_q)_{ji}$, allows to introduce the trace operation for K matrices
which is
invariant under the quantum group coaction (10) up to $\qdet T$ factor
$$ \qtr K\equiv\tr\eps_q^t K=\qtr TKT^t/\qdet T, \qquad
   T\in F_q(GL(2))                                   \eqno(16)$$

Now we can prove that $c_1$ and $c_2$ defined by (9) lie in the center of
$\A$. Note that, by definition, $c_1=\qtr K$. Take $\qtr$ of the relation
(2) with respect to the first space in $\C^2\otimes \C^2$. Then, using the
fact that the R matrix, as a solution to the YBE, is a representation of
the algebra $\F$ (4) having the quantum determinant $\qdet R=q$, and
the identities (15), (16) one obtains from the left-hand-side of (2) the
equality
$$ \qtr^{(1)}(R\one KR^{t_1})\two K=qc_1(K)\two K                      $$
and from the r.h.s. of (2) respectively
$$ \two K\qtr^{(1)}(R^{t_1}\one KR)=\two Kc_1(K)q                        $$
which establishes the commutativity of $c_1(K)$ with the generators $K$ of
$\A$.

In order to obtain the expression for $c_2$ analogous to (13) multiply
the RE (2) from the left by by the permutation operator $\P$
$$ \hat R\one KR^{t_1}\two K=\one KR^{t_1}\two K\hat R                    $$
and then by the rank one projector $P_-$. Taking into account (14) one gets
$$ P_-\one KR^{t_1}\two K=P_-\one KR^{t_1}\two KP_-=c_2(K)P_-           $$

The proof of the commutativity of $c_2(K)$ with the generators $K$ of $\A$
uses the same fusion technique as the above proof for $\qdet T$. The fact
that the quantum determinants of $R$ and $R^{t_1}$ are numbers
$$ P_-^{(12)}R_{32}R_{31}=qP_-^{(12)},\qquad
   P_-^{(12)}R_{31}^{t_3}R_{32}^{t_3}=qP_-^{(12)}                $$
again plays the crucial role.

Rewriting $q^2c_2(K)\three K$ as
$$q^2P^{(12)}_-\one KR^{t_1}\two K\three K=
     P^{(12)}_-\one KR_{12}^{t_1}\two K
   R_{31}^{t_3}R_{32}^{t_3}\three KR_{32}R_{31}                       $$
using then the YBE for $R$ and $R^{t_1}$ and RE for several times
one arrives finally to
$$q^2\three KP^{(12)}_-\one KR^{t_1}\two K=q^2\three Kc_2(K)
                                                                         $$
which proves the assertion.

Note that the central elements are transformed homogeneously under
the quantum group coaction (10)
$$\phi:c_1(K)\rightarrow\qdet Tc_1(K),\qquad
       c_2(K)\rightarrow(\qdet T)^2c_2(K).                      \eqno(17)$$

The transformation law for $c_1(K)$ follows from (16). The corresponding
formula for $c_2(K)$ follows from (15), (16) and the identity
$$\qtr K\eps_qK=(1+q^2)c_2(K)-q(c_1(K))^2                                $$
Furthermore
$$
 K\eps_qK=c_2(K)\eps_q-qc_1(K)K.  $$

The formula (17) and the following inversion formula for K matrix
$$ K^{-1}={1\over c_2}\pmatrix{\d&-\b+\om\g\cr
                   -q^2\g&\a}                                           $$
allow to think of $c_2$ as the quantum deteminant of $K$.

We have no proof that for generic $q$ the center of $\A$ is generated by
$c_1$, $c_2$ though it is highly plausible conjecture taking into account
the analogous results for the quantum group $\F$.

4. Let us discuss now the real forms of $\A$. It is well known that for
$F_q(GL(2))$ there are three real forms [5].
 For $F_q(U(2))$ the parameter $q$ is real $\bar q=q$ and the
$*$-antiinvolution is given by
$$\{a^*,b^*,c^*,d^*\}=\{d,-qc,-q^{-1}b,a\} $$

For $F_q(U(1,1))$ again $\bar q=q$ and
$$\{a^*,b^*,c^*,d^*\}=\{d,qc,q^{-1}b,a\} $$

For $F_q(GL(2,\R))$ the parameter $q$ is unitary $\bar q=1/q$ and
$$\{a^*,b^*,c^*,d^*\}=\{a,b,c,d\} $$

The corresponding real forms of $\A$ are:

$$ F_q(U(2)):\qquad \{\a^*,\b^*,\g^*,\d^*\}=\{q\d,-\b,-\g,q^{-1}\a\} $$

$$ F_q(U(1,1)):\qquad \{\a^*,\b^*,\g^*,\d^*\}=\{q\d,\b,\g,q^{-1}\a\} $$

$$ F_q(GL(2,\R):\qquad \{\a^*,\b^*,\g^*,\d^*\}=\{\a,\g+qc_1,q\g,\d\} $$

It is easy to verify that these real forms of $\A$ are consistent with
those of $\F$ that is $\phi(K_{ij}^*)=\phi(K_{ij})^*$.

5. The invariance of the ideal $c_1=0$ under the coaction $\phi$ follows
immediately from (17). The resulting quotient algebra is generated by three
generators $\a$, $\g$, $\d$ ($\b=q\g$) and the relations
$$ \a\g=q^2\g\a,\qquad \g\d=q^2\d\g,\qquad [\a,\d]=q(q^2-q^{-2})\g^2 $$

The algebra is isomorphic to the quantum homogeneous space for $\F_q(SL(2))$.
It is also isomorphic to the subalgebra of $\F$ generated by the elements
of the matrix $TT^t$ [5] (see [18] for the $F_q(GL(n))$ case).

\section{Representations}

The representation theory for the algebra $\A$ is a topic deserving special
investigation. We restrict ourselves with few remarks.

There are two one-dimensional representations of $\A$:

$$ K^{(0)}=\eps_q=\pmatrix{0&1\cr -q&0}, \qquad
   K^{(1)}=\pmatrix{\lambda&\mu\cr 0&\nu}                       \eqno(18)$$

As shown in the next section, under the additional condition of $\g$
being invertible the quotient algebra $\A/(c_2=1)$ is isomorphic to the
quantum algebra $\Uq$.
Hence, the irreducible representations of $\Uq$ are translated into
those of $\A$.

Another way to construct representations of $\A$ is to use the coaction
$\phi$ (10)
of $\F$. Any pair of representations $\pi$ of $\A$ and $\rho$ of $\F$
$$ \pi:\A\rightarrow{\rm End}(V),\qquad \rho:\F\rightarrow{\rm End}(W) $$
gives rise to another representation of $\A$
$$ (\rho\otimes\pi)\circ\phi:\A\rightarrow{\rm End}(W\otimes V)     $$

It is an open question whether this construction generates new irreducible
representation of $\A$ provided $\pi$ and $\rho$ are irreducible.
It is also unknown if any irreducible representation of $\A$ can be
decomposed into $\pi$ and $\rho$.

Representations of comodule-algebras related to the reflection equations with
R matrices corresponding to other quantum (super)groups such as
$F_q(GL(m|n))$ were studied in [19].

\section{Variants of RE and related algebras}

1. Consider the algebra $\X$ defined by the generators $x$, $y$, $u$, $v$
and by the relations
$$ \eqalign{xy&=qyx,\cr uv&=qvu,} \qquad
   \eqalign{xu&=qux,\cr yu&=uy,} \qquad
   \eqalign{xv&=vx+\om uy,\cr yv&=qvy,} $$
Introducing the column $X$ and row $Y$
$$ X=\pmatrix{x\cr y},\qquad Y=\pmatrix{u&v}  $$
one can represent the above commutation relations as the exchange algebras
[7, 8, 11]
$$ R\one X\two X=q\two X\one X, \qquad
    \two Y\one YR=q\one Y\two Y                              \eqno(19)$$

$$ \one YR^{t_1}\two X=\two X\one Y,\qquad
   \two YR^{t_1}\one X=\one X\two Y                         \eqno(20)$$

The algebra $\X$ has a central element $\zeta=Y\eps_q X=uy-qvx$.
The subalgebras of $\X$ spanned by $X$ or $Y$ are isomorphic to the ``quantum
plane'' [13].

By virtue of the relations (19), (20) the matrix
$$ K=XY=\pmatrix{xu&xv\cr yu&yv} $$
satisfies RE (2) the above formula giving thus a homomorphism $\A\rightarrow
\X$. Note that for the obtained realization of $\A$ one has $c_2(K)=0$ and
$K\eps_qK=\zeta K=-qc_1(K)K$.

2. In the papers [6, 7] another version of RE was obtained
$$ R\one MR^{-1}\two M=\two M \tilde R^{-1}\one M\tilde R, \qquad
   \tilde R\equiv\P R\P, \qquad
   M=\pmatrix{\xi&\eta\cr \theta&\tau}           \eqno(21)$$

The corresponding algebra $\B$ is an $F_q$-comodule algebra with the coaction
$$ \psi:\B\rightarrow\F\otimes\B,\qquad \psi(M)=TMT^{-1}              $$

In the case of $\F=F_q(GL(2))$ the $\F$-comodule algebra $\B$
is isomorphic to $\A$ because of the relation (15). The isomorphism is given
by the formula
$$  K=\pmatrix{\a&\b\cr \g&\d}=
      \pmatrix{-q\eta&\xi\cr -q\tau&\theta}=M\eps_q                $$
which implies the relations between the central elements
$$ \eqalign{c_1(K)&=\b-q\g=\xi+q^2\tau\equiv z_1(M) \cr
       c_2(K)&=\a\d-q^2\b\g=q^3(\xi\tau-q^{-2}\eta\theta)\equiv q^3z_2(M)}
                                                                    $$
There is a characteristic equation for the matrix $M$ also.

Under this isomorphism the q-trace (16) maps into $\tr DM$,
$D={\rm diag}(q^{-1},q)$ introduced in [5], see also [10].

The algebra $\B$ is connected closely to the quantum algebra $\Uq$ [6,7].
Namely, the matrix $M$ can be realized in terms of $L^\pm$ (12)
$$ M=s(L^+)L^-                                               \eqno(22)$$
where  $s$ is the antipode in $\Uq$: $s(H)=-H$, $s(X_\pm)=-q^{\mp 1}X_\pm$.

The formula  (22) describes
the algebra homomorphism $\chi:\B\rightarrow\Uq$
$$ \chi(M)=\pmatrix{q^{-2H}+q\om^2X_-X_+ & -\om qX_-q^H \cr
                 -\om q^HX_+ & q^{2H} }                           $$
for which $z_1(M)$ is proportional to the well known Casimir operator for
$\Uq$ and $z_2(M)=1$ [10].

It  is easy to see that the inversion of the homomorphism $\chi$
needs invertibility of the element $\tau$ (which is not the case for the
one-dimensional representation $M=K^{(1)}\eps_q$, see (18)). If the element
$\tau$ is supposed to be invertible one can introduce in $\B$ a coproduct
induced from $\Uq$ [7].

3. We conclude this section with describing the classical limit of $\A$.
Let $q=\e^h$. As $h\rightarrow 0$ or $q\rightarrow 1$ the algebra $\A$
becomes commutative, its commutator giving rise to the Poisson bracket
$[\, ,\,]=-h\{\, ,\,\}$. The commutation relations (2) are transformed into
the Poisson brackets relations
$$ \{\one K,\two K\}=[r,\one K\two K]+\one Kr^{t_1}\two K
                                     -\two Kr^{t_1}\one K           \eqno(23)$$
where $r$ is the classical r matrix $R=1+hr+O(h^2)$
$$ r=\pmatrix{1&&&\cr
              &&&\cr
              &2&&\cr
              &&&1}, \qquad
     r^{t_1}=\pmatrix{1&&&2\cr
              &&&\cr
              &&&\cr
              &&&1}, \qquad                                        $$

Fixing values of the central functions $c_1=\b-\g$ and $c_2=\det K=\a\d-\b\g$
one obtains the foliation of the 4-dimensional space spanned by $\a\b\g\d$
into 2-dimensional manifolds on which the Poisson bracket (23) is
nondegenerate.

\section{Generalizations}

1. Let us discuss now the fusion procedure for RE. It does not differ much
from the case of the YBE [5, 16]. The peculiarities of the RE case are well
seen in the simplest spin-1 case. The R matrix in (4) $R\in\C^2\otimes\C^2$
intertwines two spin-1/2 corepresentations of $F_q(GL(2))$. The R matrix in
$\C^2\otimes\C^3$ is
$$ R_{1(2)}=P_+R_{12^\prime}R_{12}=P_+R_{12^\prime}R_{12}P_+        $$
where $P_+$ is the rank 3 projector (14) from $\C^2\otimes\C^2$ onto $\C^3$
and the corresponding spaces are labelled respectively by the indices $2$,
$2^\prime$ and $(2)$.
We shall use also
$$ R_{1(2)}^{t_1}=P_+R_{12}^{t_1}R_{12^\prime}^{t_1}P_+     $$

The $3\times 3$ matrix $J$
$$ J=P_+\two K R_{22^\prime}^{t_1}\twoprime K P_+               $$
satisfy the RE in $\C^2\otimes\C^3$
$$ R_{1(2)}\one K R_{1(2)}^{t_1}\twopar J=
   \twopar J R_{1(2)}^{t_1}\one K R_{1(2)}                     $$
and also the RE in $\C^3\otimes\C^3$
$$ R_{(1)(2)}\onepar J R_{(1)(2)}^{t_1}\twopar J=
   \twopar J R_{(1)(2)}^{t_1}\onepar J R_{(1)(2)}              $$

The entries of the matrix $J$ generate a subalgebra $\A^\prime$ of $\A$ which
is also an $\F$-coideal that is $\phi:\A^\prime\rightarrow\F\otimes\A^\prime$.
The analogous fusion procedure for the spectal
parameter dependent RE (1) was developed in [20].

There is no doubt that, being properly generalized, the above described
fusion procedure should be able to produce solutions to RE in
matrices of any dimension corresponding
to higher finite-dimensional irreducible corepresentations of $F_q(GL(2))$.
At least for the one-dimensional representations of $\A$ (18) the
expression for the ``universal K matrix'' can be written explicitely
(see [21] for $K^{(1)}$ and [12, 19] for $K^{(0)}$).

2. Throughout this paper we considered the algebra $\A$ defined by the
relations
(2) corresponding to the R matrix (5) of the quantum group $F_q(GL(2))$.
However, the proof of the comodule algebra property does not use anything
besides the commutation relations (4) and the symmetries (11) of the R matrix.
Therefore, to any R matrix satisfying (11) and the related quantum group
$\FR$ defined by relations (4) there corresponds some
$\FR$-comodule-algebra $\AR$ defined by relations (2).

The conditions (11), however, are not essential and can be disposed of.
Let us extend the algebra $\FR$ introducing another matrix of generators
$S$ in addition to $T$ and imposing the commutation relations
$$ \eqalign{R^{(1)}\one T \two T&=\two T\one T R^{(1)},\cr
            \one TR^{(3)}\two S&=\two SR^{(3)}\one T,} \qquad
   \eqalign{\one SR^{(2)}\two T&=\two TR^{(2)}\one S, \cr
            R^{(4)}\one S \two S&=\two S\one SR^{(4)}}
                                                            \eqno(24) $$
parametrized by four R matrices. It is easy to verify that if $K$
obeys the relation
$$ R^{(1)}\one KR^{(2)}\two K=\two KR^{(3)}\one KR^{(4)}  \eqno(25)$$
then the matrix $TKS$ obeys the same relation provided matrix elements of
$K$ commute with those of $T$ and $S$.

Suppose now that $S=\sigma(T)$ where $\sigma$ is an
antiautomorphism of matrix algebra that is $\sigma(xy)=\sigma(y)\sigma(x)$
for any number matrices $x$, $y$. Then all the R matrices in (25) can be
expressed in terms of the matrix $R$ defining the quantum group (4).
The resulting algebra is an
$\FR$-comodule-algebra with the coaction $K\rightarrow TK\sigma(T)$.

Note that there exists a remarkable ambiguity in the choice of the R matrices.
Namely, one can make the substitution $R\rightarrow\P R^{-1}\P$ independently
in any of the matrices $R^{(i)}$ which does not affect the commutation
relations (24), \cf the remark immediately after formula (5).  Consequentely,
there exist at least $2^4=16$ {\it a priori} nonequivalent algebras sharing
the same comodule structure. The interrelations of these algebras will be
described in a separate paper.

In the case of $\sigma(T)=T^t$ the R matrices are
$$ \eqalign{R^{(1)}&=R \cr
            R^{(2)}&=R^{t_1} \cr
            R^{(3)}&=(\tilde R)^{t_2} \cr
            R^{(4)}&=\tilde R^t }
   \eqalign{ {\rm or\ } & \tilde R^{-1} \cr
              {\rm or\ } & (\tilde R^{-1})^{t_1} \cr
              {\rm or\ } & (R^{-1})^{t_2} \cr
              {\rm or\ } & (R^{-1})^t \cr }                        $$
where, as in (21), the notation $\tilde R\equiv\P R \P$ is used.
Choosing the first option for every $R^{(i)}$ one obtains from (25) the
equation [19]
$$ R\one KR^{t_1}\two K=\two K\tilde R^{t_2}\one K \tilde R^t    \eqno(26)$$
coinciding with the RE (2) if $R$ has the symmetries (11).

Another example is provided by $\sigma(T)=T^{-1}$.
In this case
$$ \eqalign{R^{(1)}&=R \cr
            R^{(2)}&=R^{-1} \cr
            R^{(3)}&=\tilde R^{-1} \cr
            R^{(4)}&=\tilde R}
   \eqalign{ {\rm or\ } & \tilde R^{-1} \cr
              {\rm or\ } & \tilde R \cr
              {\rm or\ } & R \cr
              {\rm or\ } & R^{-1} \cr }                             $$

The choice of the first option for every $R^{(i)}$ leads to the algebra $\B$
described in the section 5.2.


3. We conclude our discussion of RE with a remark concerning the very form of
the equation. Note that formula (10) for the coaction of $\F$ is equivalent
to
$$ \phi:K^{i_1i_2}\rightarrow T^{i_1}_{j_1}T^{i_2}_{j_2}K^{j_1j_2} \eqno(27)$$
where the indices are shown explicitely. Here we use the commutativity of
$T$ and $K$ and assume summation over repeated indices.

The formula (27) suggests that $K$ can be thought of as a bivector
(contravariant tensor of the second order). In a compact form one can write
(27) as
$$ \phi: K_{12}\rightarrow \one T\two T\cdot  K_{12}               \eqno(28)$$
(the notation is self-evident).

Analogously,  the equation (26) can be rewritten as
$$R^{i_1i_2}_{j_1j_2}K^{j_1j^\prime_1}
  R^{i^\prime_1j_2}_{j^\prime_1k_2}K^{k_2i^\prime_2}=
  K^{i_2j^\prime_2}R^{k^\prime_2i_1}_{j^\prime_2j_1}
  K^{j_1j^\prime_1}R^{i^\prime_2i^\prime_1}_{k^\prime_2j^\prime_1} $$
or, using the same notation as in (28)
$$ R_{12}R_{1^\prime2}\cdot K_{11^\prime}K_{22^\prime}=
   \tilde R_{1^\prime2^\prime}\tilde R_{12^\prime}\cdot
   K_{22^\prime}K_{11^\prime}                                       \eqno(29)$$

Moving two R matrices from the right hand side of (29) to the left
hand side one obtains
$$ \tilde R^{-1}_{12^\prime}\tilde R^{-1}_{1^\prime 2^\prime}
    R_{12}R_{1^\prime 2}\cdot K_{11^\prime}K_{22^\prime}=
    K_{22^\prime}K_{11^\prime}                                    \eqno(30)$$

Combining the two indices in $K^{i_1i_2}$ into one composite index and the
four R matrices in (30) into one composite R matrix we observe that the
equation (30) describes a sort of exchange algebra (19).

The case $\sigma(T)=T^{-1}$ leads in the same way to a mixed tensor
having one upper and one lower index.

This observation opens  a road to a far-going
generalization of the algebra $\A$. Acting in the same spirit one can
introduce algebras corresponding to the
quantum tensors with any number of upper and lower indices.
The papers [13, 22] where the quantum
multilinear algebras are studied might be
of relevance in performing this program.
We hope to touch this subject in a separate article.

\ack
We are grateful to our colleagues at YITP and RIMS of Kyoto University for
warm hospitality and useful discussions.

\references
\numrefjl{1}{Sklyanin E 1988}{\JPA}{21}{2375}
\numrefjl{2}{Cherednik I 1984}{Theor.\ Math.\ Phys.}{61}{55}
\numrefjl{3}{Reshetikhin N and Semenov-Tian-Shansky M 1990}{Lett.\ Math.\
Phys.}{19}{133}
\numrefjl{4}{Kulish P and Sklyanin E 1991}{\JPA}{24}{L435}
\numrefjl{5}{Faddeev L, Reshetikhin N and Takhtajan L 1989}
{Algebra i Analiz}{1}{178}
\numrefjl{}{Takhtajan L 1990}{Lecture Notes Phys.}{370}{3}
\numrefjl{6}{Moore G and Reshetikhin N 1989}{Nucl.\ Phys.}{B328}{557}
\numrefjl{7}{Alekseev A, Faddeev L and Semenov-Tian-Shansky M 1992}
{Lecture Notes Math.}{1510}{148}
\numrefjl{}{Alekseev A and Faddeev L 1991}{Commun.\ Math.\ Phys.}{141}{413}
\numrefjl{8}{Babelon O 1991}{Commun.\ Math.\ Phys.}{139}{619}
\numrefjl{9}{Freidel L and Maillet J 1991}{\PL}{B262}{278; {\bf B263} 403}
\numrefbk{10}{Zumino B 1991}{Introduction to the differential geometry
of quantum groups.}{Preprint UCB-PTH-62/91}
\numrefjl{11}{Kulish P 1991}{\PL}{A161}{50}
\numrefbk{12}{Kulish P 1991}{Quantum groups and quantum algebras as
symmetries of dynamical systems.}{Preprint YITP/K-959}
\numrefbk{13}{Manin Yu 1991}{Topics in noncommutative geometry.}{Princeton
Univ. Press}
\numrefjl{14}{Rosso M 1988}{Commun.\ Math.\ Phys.}{117}{581}
\numrefjl{15}{Majid S 1990}{Int.\ J.\ Mod.\ Phys.}{A5}{1}
\numrefjl{}{\dash 1991}{J.\ Math.\ Phys.}{32}{3246}
\numrefjl{16}{Kulish P, Reshetikhin N and Sklyanin E 1981}{Lett.\ Math.\ Phys.}
{5}{393}
\numrefjl{17}{Jimbo M 1986}{Lett.\ Math. Phys.}{11}{247}
\numrefbk{18}{Jing N and Yamada H 1990}{Polyn\^omes zonaux pour le groupe
lin\'eaire g\'en\'eral quantique.}{Preprint IAS}
\numrefbk{}{Noumi M 1992}{Macdonald's symmetric polynomials as zonal spherical
functions on some quantum homogeneous spaces}{Preprint UT-Komaba}
\numrefbk{19}{Kulish P, Sasaki R and Schwiebert C 1992}{Constant solutions
of reflection equations.}{Preprint YITP/U-92-07}
\numrefbk{20}{Mezincescu L and Nepomechie R 1991}{Fusion procedure for open
chains.}{Preprint CERN-TH.\ 6152/91}
\numrefjl{21}{Cremmer E and Gervais J-L 1992}{Commun.\ Math.\ Phys.}{144}{279}
\numrefbk{22}{Hashimoto M and Hayashi T 1991}{Quantum multilinear
algebra.}{Preprint Nagoya Univ.}
\numrefjl{}{Gurevich D and Rubtsov V 1992}{Lecture Notes Math.}{1510}{47}

\bye